\documentclass[letter]{aa}
\usepackage[T1]{fontenc}
\usepackage[varg]{txfonts}
\usepackage[utf8]{inputenc}
\usepackage{graphicx,color}
\usepackage{natbib}
\usepackage{epstopdf}

\usepackage[normalem]{ulem}

\begin{document}

\def\vecv{\mbox{\boldmath $v$}}
\def\vecB{\mbox{\boldmath $B$}}
\def\vecq{\mbox{\boldmath $q$}}
\def\vecQ{\mbox{\boldmath $Q$}}

\title{Magnitude and direction of the local interstellar magnetic field inferred from Voyager 1 and 2 interstellar data and global heliospheric model}
\titlerunning{IsMF inferred from Voyagers and global heliosphere model}
\authorrunning{Izmodenov et al.}
\author{Vladislav V. Izmodenov\inst{1,2,3}
        \and Dmitry B. Alexashov\inst{1,3}
        }
\offprints{V. Izmodenov, \email{izmod@iki.rssi.ru}}
        \institute{Space Research Institute (IKI) of Russian Academy of Sciences
                        \and Lomonosov Moscow State University
                \and Institute for Problems in Mechanics Russian Academy of Sciences}
        
        \date{Received 5 November 2019 / Accepted 26 December 2019}
        
        \abstract{In this Letter, we provide constraints on the direction and magnitude of the pristine (i.e., unperturbed by the interaction with the Sun) local interstellar magnetic field. The constraints are based on analysis of the interstellar magnetic field components at the heliopause measured by magnetometer instruments on board Voyager 1 and 2 spacecraft. The analysis was performed with the help of our kinetic-magnetohydrodynamical (MHD) model of the global heliosphere. The model shows that the solar-induced disturbances of the interstellar magnetic field are extended relatively far from the Sun up to 400-500 AU. The field is draped around the heliopause and compressed. By comparison of the model results with Voyager data we found that the model provides results comparable with the data for the interstellar magnetic field of 
 $B_{LISM}$ = 3.7-3.8 $\mu$G in magnitude and directed towards  $\approx$125$^\circ$ in longitude, and  $\approx$37$^\circ$ in latitude in the heliographic inertial (HGI) coordinate system.}

        
\maketitle

\section{Introduction}
\label{intro}

Until 2012, the local interstellar magnetic field (IsMF) could only be determined indirectly.
 \cite{Lallement_2005, Lallement_2010} analyzed Solar and Heliospheric Observatory (SOHO)/ Solar Wind ANisotropy (SWAN) instrument data on backscattered Lyman-$\alpha$ emission and concluded that the direction of the interstellar neutral hydrogen flow in the heliosphere is deflected by $\sim$ 4$^\circ$ from the direction of the pristine (i.e., undisturbed by the interaction with the Sun) local interstellar gas flow. 
The deflection is due to distortion of the global shape of the heliosphere under the action of the inclined interstellar magnetic field.
The interstellar H atoms pass through the heliospheric interface and interact with the plasma component by charge exchange. Imprints of the asymmetry of the heliospheric plasma interface in the distribution of the interstellar H atom component appear as an observed deflection of the H atom flow as compared with the interstellar helium.  
\cite{Lallement_2005} suggested that the plane determined by the velocity vectors of interstellar helium and hydrogen inside the heliosphere coincides with the plane determined by the vectors of velocity,  ${\bf V}_{LISM}$, and magnetic field, ${\bf B}_{LISM}$, of the pristine local interstellar medium. This plane is often referred to as the (BV)-plane or hydrogen deflection  plane (HDP).  \cite{Izmodenov_2005, Izmodenov_2009} applied their 3D kinetic-magnetohydrodynamical (MHD) models of the solar wind (SW) interaction with the local interstellar medium
(LISM) and confirmed the value of deflection angle observed by \cite{Lallement_2005}. Later, \cite{Katushkina_2015} studied the deflections in detail and confirmed these previous conclusions.

The second way to put constraints on the LISM magnetic field is the asymmetry of heliospheric termination shock (TS) in Voyager 1 and 2 directions. Indeed, Voyager 1 crossed the TS at 94 AU in 2004 while Voyager 2 at 84 AU in 2007. It was  surprising that Voyager 2 crossed the TS
closer by 10 AU because the trajectory of  Voyager 2 was further away from the direction of interstellar flow  as compared to Voyager 1. Therefore, prior to the crossing it was expected that the TS
distance in Voyager 2 direction was larger than 94 AU.
However, the solar wind dynamic pressure decreases in the period from 2004 to 2007, and so in this time period the TS moved toward the Sun  (e.g., \cite{Izmodenov_2008}). Nevertheless, these time variations can explain only a small part of the TS asymmetry. The main reason for the asymmetry of the global heliosphere, and in particular the termination shock, is the inclined interstellar magnetic field. The strength of the magnetic field should be large enough to create the asymmetry. \cite{izmod_alexash_2015} employed the latest version of a 3D kinetic-MHD model of the global heliosphere and concluded that the model with an interstellar magnetic field of $B_{LISM} \approx 4.4$ $\mu$Gauss in magnitude  and (163.5$^\circ$, 18$^\circ$) in direction (in HGI coordinate system) provided termination shock distances in Voyager 1 and 2 directions that are comparable with the distances of actual crossings.
However, the distance to the heliopause obtained in the frame of this model was $\sim$164 AU, which is very far
from the actual distance  of the heliopause crossing
by Voyager 1 of 122 AU.
 It was suggested that this discrepancy can be explained by a dissipation mechanism in the inner heliosheath (e.g., \cite{Izmodenov_2014}; see also discussions in \cite{izmod_alexash_2015, Izmodenov_2018}).

Another way to determine the magnitude and direction of the pristine interstellar field is to analyze the ribbon of enhanced fluxes of heliospheric 
energetic neutral atoms (ENAs) that was discovered by 
Interstellar Boundary Explorer (IBEX) \citep{McComas_2009}. Soon after this discovery, it was realized \citep{McComas_2009, Heerikhuisen_2010, Chalov_2010} that the enhancement of the fluxes could be in the directions where the radial component of the interstellar magnetic field around the heliopause is close to zero.
Zirstein et al. (2016) modeled the ENA ribbon using a model of the global heliosphere, and after comparison with IBEX data concluded that the interstellar magnetic field is
$\approx$3 $\mu$G in magnitude and  has the following direction in  HGI coordinates: (longitude = 147.5$^{\circ}$, latitude = 31$^{\circ}$).

Since the crossings of the heliopause by Voyagers in 2012 and 2018, their magnetometers are providing us unique measurements of the interstellar magnetic field. However, 
the measured magnetic field is not  the pristine interstellar field because the interstellar plasma and magnetic field are strongly disturbed by an interaction with the solar wind. 

The goal of this study is to use constraints provided by the magnetic field components measured by Voyager magnetometers and the locations of the heliopause in Voyager 1 and 2 directions to obtain a new independent estimate of the magnitude and direction of the pristine interstellar magnetic field.
To perform the analysis we use the time-dependent version of the model by Izmodenov and Alexashov (2015).  Details on the model are given in Appendix A.

\section{Preliminary qualitative consideration}

 The magnetometers onboard Voyager 1 and 2  measured, for the first time, the interstellar magnetic field after they crossed the heliopause. A detailed comparison of the three magnetic field components  measured by Voyagers 1 and 2 with those obtained in the model are provided in the following section. In this section we consider the consequences of two observational phenomena: 1) the radial component 
of the magnetic field $B_R$ is positive in Voyager 1 direction, and negative but close to zero in Voyager 2 direction; and 2) the  $B_T$ component of the magnetic field in
radial tangential normal (RTN) coordinate system (see, e.g., \cite{Hapgood}) is negative in both directions.
These facts provide restrictions on the local shape of the heliopause in directions of crossings and also on the topology of the interstellar magnetic field around the heliopause (\cite{burlaga_2018, burlaga_2019, burlaga_2019b}; see also Figure 3).

\begin{figure}[t!]
        \includegraphics[width=0.5\textwidth,clip=]{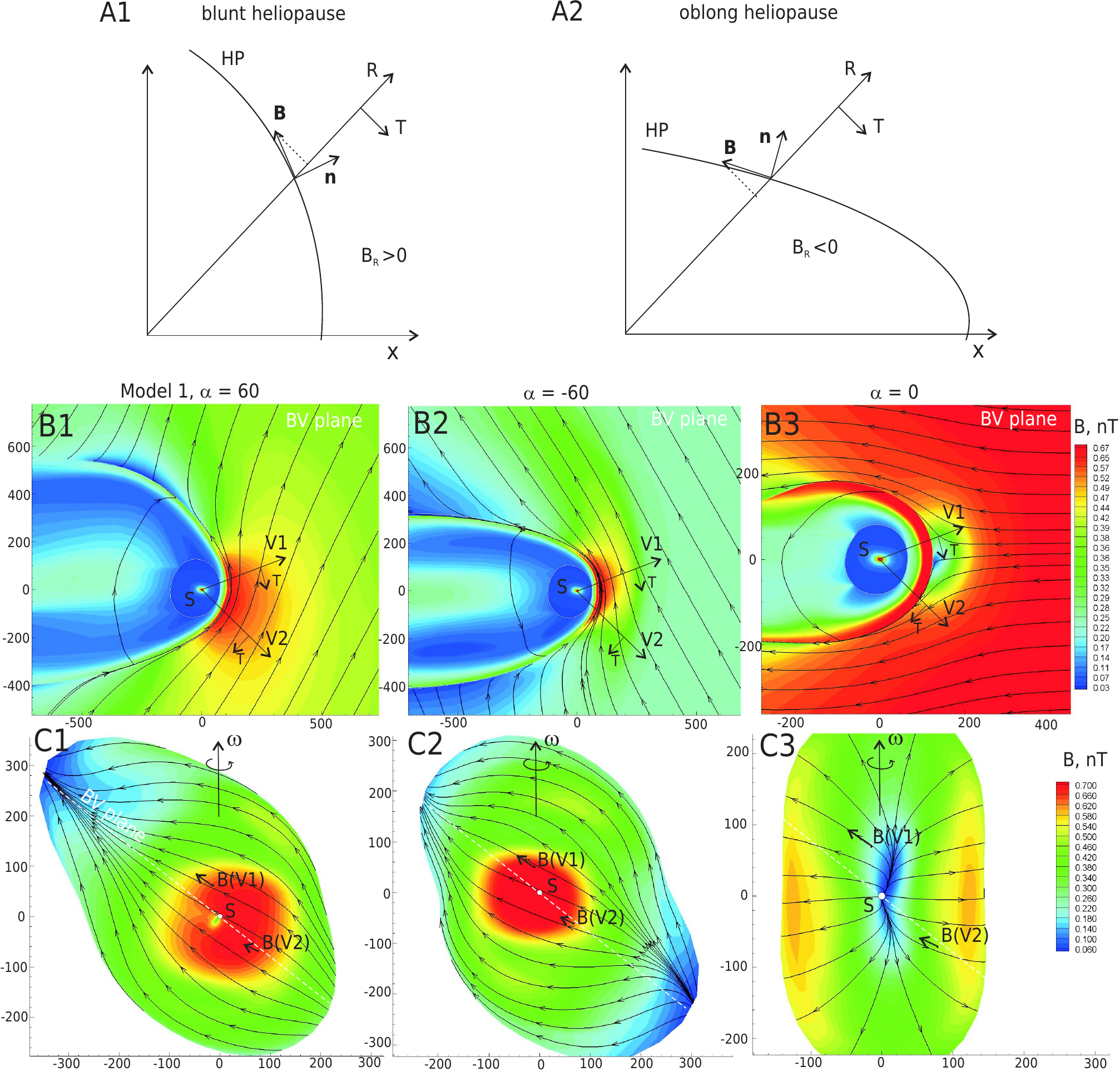}
        \caption{{\bf Panel A}: Sketch of the blunt (A1) and oblong (A2) shape of the heliopause.
                {\bf Panel B}: Magnetic field lines and magnitudes of the magnetic field in the plane determined by $\mathbf{V}_{LISM}$ and $\mathbf{B}_{LISM}$ vectors. {\bf Panel B1}: a model with the magnetic  field critical point above Voyager 1 (Model 1 hereafter);
                {\bf Panel B2}: a model with the magnetic  field critical point ($\mathbf{B} =0$) below Voyager 2; {\bf Panel B3}: a model with the critical point located in between Voyager 1 and 2 (B3).  {\bf Panel C}: Projections of magnetic field lines on the heliopause. Panels C1, C2, and C3 correspond to the same models as panels B1, B2, and B3, respectively. 
        }\label{fig1}
\end{figure}

If we assume an ideal nondissipative approach (as in the model) then the heliopause is a tangential discontinuity at which $B_n = 0$, where $B_n$ is the projection of the magnetic field vector to the normal of the heliopause surface. Therefore,   
 the  magnetic field vector should be  parallel to the surface of the heliopause.
In this case, the sign of the $B_R$ component depends on the local shape of the heliopause. If the heliopause is locally spherical then $B_R =0$. If the projection of the normal $\mathbf{n}$ to the heliopause on the X-axis (toward upwind) is larger than the X-axis projection of the unit radius vector then heliopause has a blunt shape (Figure \ref{fig1}, panel A1). Otherwise, the shape is oblong (Figure \ref{fig1}, panel A2). The sketches in panels A1 and A2  also show that in the case of negative 
$B_T$:  $B_R >0$  for the blunt shape, and  $B_R <0$  for the oblong case. 
If the direction of the magnetic field is opposite (not shown in the sketches),  then  $B_R <0$  for the blunt shape, and  $B_R >0$  for the oblong case.
In the direction of Voyager 1, $B_T <0$ and   $B_R >0$, and therefore the heliopause has to have a blunt shape in this direction.  
In the direction of Voyager 2, $B_T <0$ and   $B_R <0$, and therefore the heliopause has to have an oblong shape in this direction.   

 The fact that $B_T$ is negative in both Voyager 1 and Voyager 2 directions also   means that the critical point of the magnetic field (i.e., the point of $\mathbf{B} =0$) at the heliopause should be either above Voyager 1 or below Voyager 2 directions. 
 The two possible cases with  $B_T <0$ are shown in panels  B1 and B2 of Figure \ref{fig1} (see, also, panels C1, C2 for another projection). In the case of panel B1(C1) the critical point is above Voyager 1, and in the case of panel B2(C2), the critical point is below Voyager 2.
 If the critical point is located between Voyager 1 and Voyager 2 directions then the sign of $B_T$ will be different in these directions. This is illustrated in panels B3 and C3 of Figure \ref{fig1}. Panels B1-C3 present the results of model calculations; these are shown for illustrative purposes only.

In the vicinity of the magnetic field critical point, the magnetic tension stretches the heliopause out of the Sun (see, \cite{Baranov_Zaitsev_1995}).   This causes the shape of the heliopause to be locally blunt in a significant part of it close to the stagnation point.  At the same time, in the other hemisphere, where the magnetic field is almost tangential, the magnetic pressure pushes the heliopause closer to the Sun. 

Panels B1 and B2 of Figure \ref{fig1} clearly illustrate the stretching and pushing of the heliopause. In the case of panel B1 the heliopause has a blunt shape in Voyager 1 direction and an oblong shape in Voyager 2 direction. Therefore, $B_R >0$ and $B_R <0$ in Voyager 1 and 2 directions, respectively. These calculations are in the agreement with the observations of Voyager.
For the case shown in panel B2, the heliopause is oblong and  $B_R <0$  in Voyager 1 direction, while the heliopause is blunt and  $B_R > 0$ in Voyager 2 direction. Therefore, the case of panel B2 should be ruled out.

We conclude in this section that the signs of $B_R$ and $B_T$ components measured in Voyager 1 and 2 directions provide qualitative constraints on the magnetic field configurations in the vicinity of the heliopause. Namely, the configuration should be qualitatively similar to the case shown in Panel B1, that is, the critical point of the magnetic field should be above the Voyager 1 direction. The scenarios shown in panels B2 and B3 are ruled out.

\section{Results and comparison with Voyager data}

In this section, we present the results of the model with  B$_{LISM}$ = 3.75 $\mu$G and  $\alpha$=60$^{\circ}$ (referred to as Model 1), where the magnetic field belongs to the hydrogen deflection plane and $\alpha$ is the angle between vectors of the interstellar velocity and magnetic field. The direction of the interstellar magnetic field in the HGI system is $\approx$125$^\circ$ in longitude and  $\approx$37$^\circ$ in latitude.
 For comparison, the results of the model from Izmodenov and Alexashov (2015) with  B$_{LISM}$ = 4.4 $\mu$G and $\alpha$=20$^{\circ}$ are also shown and are referred to as Model 2.
We note that both models are "Voyager permitted" that is, they correspond qualitatively to the case shown in Panel B1 in Figure \ref{fig1}.

\begin{figure}[t!]
        \includegraphics[width=0.5\textwidth,clip=]{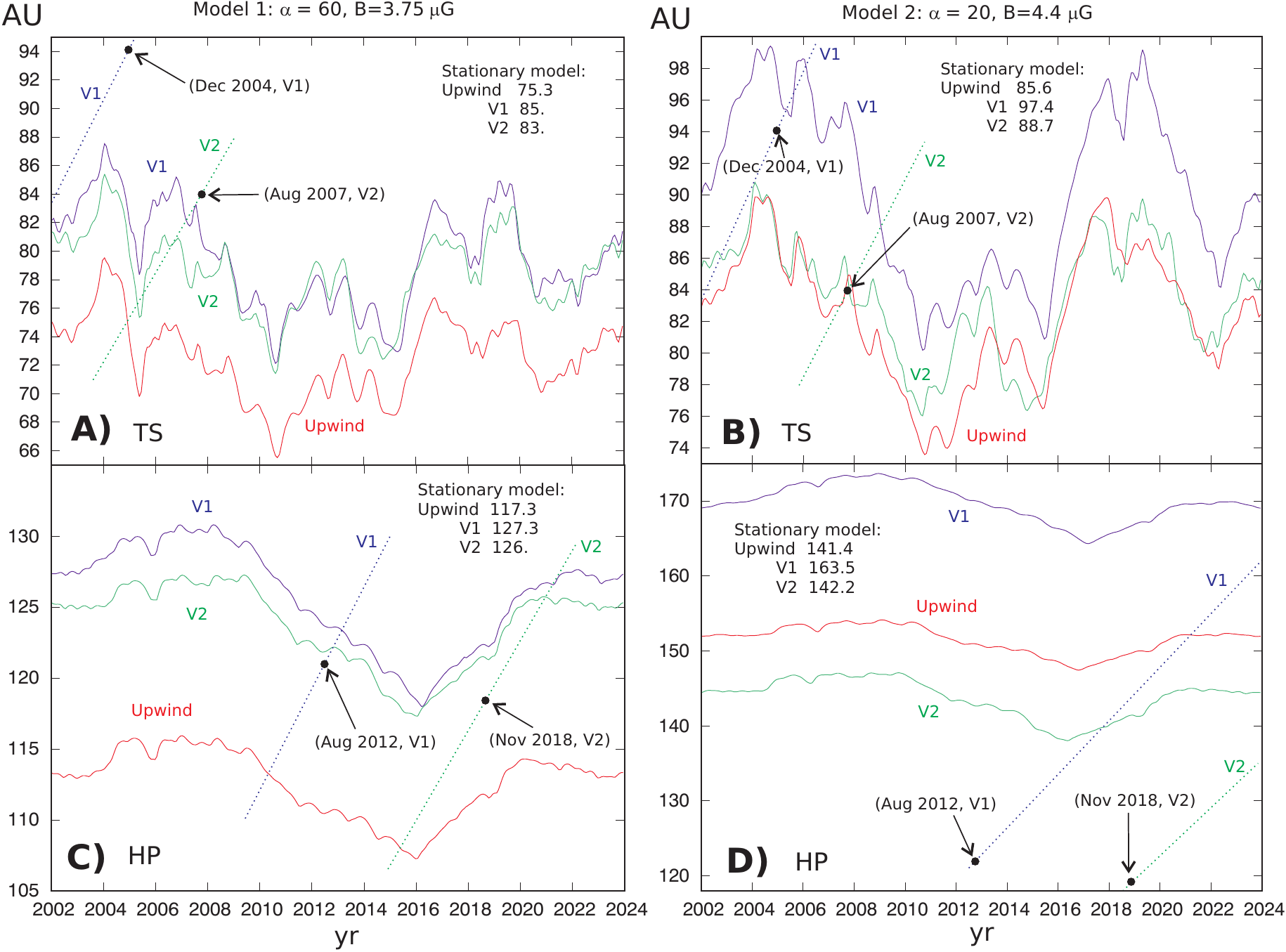}
        \caption{Heliocentric distances to the heliospheric termination shock (A and B) and heliopause (C and D) are shown for the directions of Voyager 1 (blue curves), Voyager 2 (green curves), and in the upwind direction (red curves). 
                Panels A and C correspond to Model 1, panels B and D to Model 2.
        }\label{fig2}
\end{figure}

Before comparing the model results with data obtained by the Voyager spacecraft, we present (Figure \ref{fig2}) fluctuations of the heliospheric termination shock and the heliopause with time. The termination shock fluctuates by $\sim$ 12- 15 AU from minimal to maximal distance depending on the direction that is in agreement with previous time-dependent models
\citep{Izmodenov_2005b, Izmodenov_2008}. The heliopause distances obtained by Model 1 are $\sim$123.5 AU  in August 2012 for Voyager 1 direction and $\sim$121.5 AU  in November 2018 for Voyager 2 direction. These distances do not exactly coincide with the distances of actual heliopause crossings by Voyagers, but are closer than the results obtained in the frame of Model 2, which are $\sim$ 170 AU and 141 AU, respectively. Such a dramatic differences in the distances to the heliopause and in the heliopause asymmetry demonstrates that the angle $\alpha$ between $\mathbf{V}_{LISM}$ and   $\mathbf{B}_{LISM}$ vectors is very important. The larger the angle, the closer the perpendicular component of the magnetic field pressure pushes the heliopause toward the Sun, and  the smaller the parallel component that makes the heliopause more asymmetric (for $\alpha \neq 0$).  The model results show that the heliopause moves inside from 2009 to 2016 and then moves outside. This is surely connected with the minimum of the solar wind dynamic pressure from 2007-2008 to 2015 as seen in Figure \ref{sw_1AU}.
It is interesting to note that although the excursions of the TS with time are quite similar in the two presented models, the small-scale details are different. We connect this with the difference in the global shape of the heliopause. 

The distances to the termination shock are $\sim$84 AU in September 2004, and $\sim$ 78 AU in July-August 2007. Therefore, Model 1 does not reproduce a  10 AU difference from the actual distances of Voyager 1 and 2 crossings, but reproduces a $\sim$6 AU difference, which is a significant fraction of 10 AU. The model distances to the TS are by 10 AU and 6 AU (for Voyager 1 and 2, respectively) smaller as compared with actual Voyager crossings.    

Now let us compare the magnetic field component obtained in Model 1 with the actual Voyager data. The comparison of the model results with Voyager data in the inner heliosheath is not a major aim of this study, but is discussed in Appendix B. Here we focus on the comparison of the magnetic field components in the outer heliosheath beyond the heliopause.

\begin{figure}[t!]
        \includegraphics[width=0.5\textwidth,clip=]{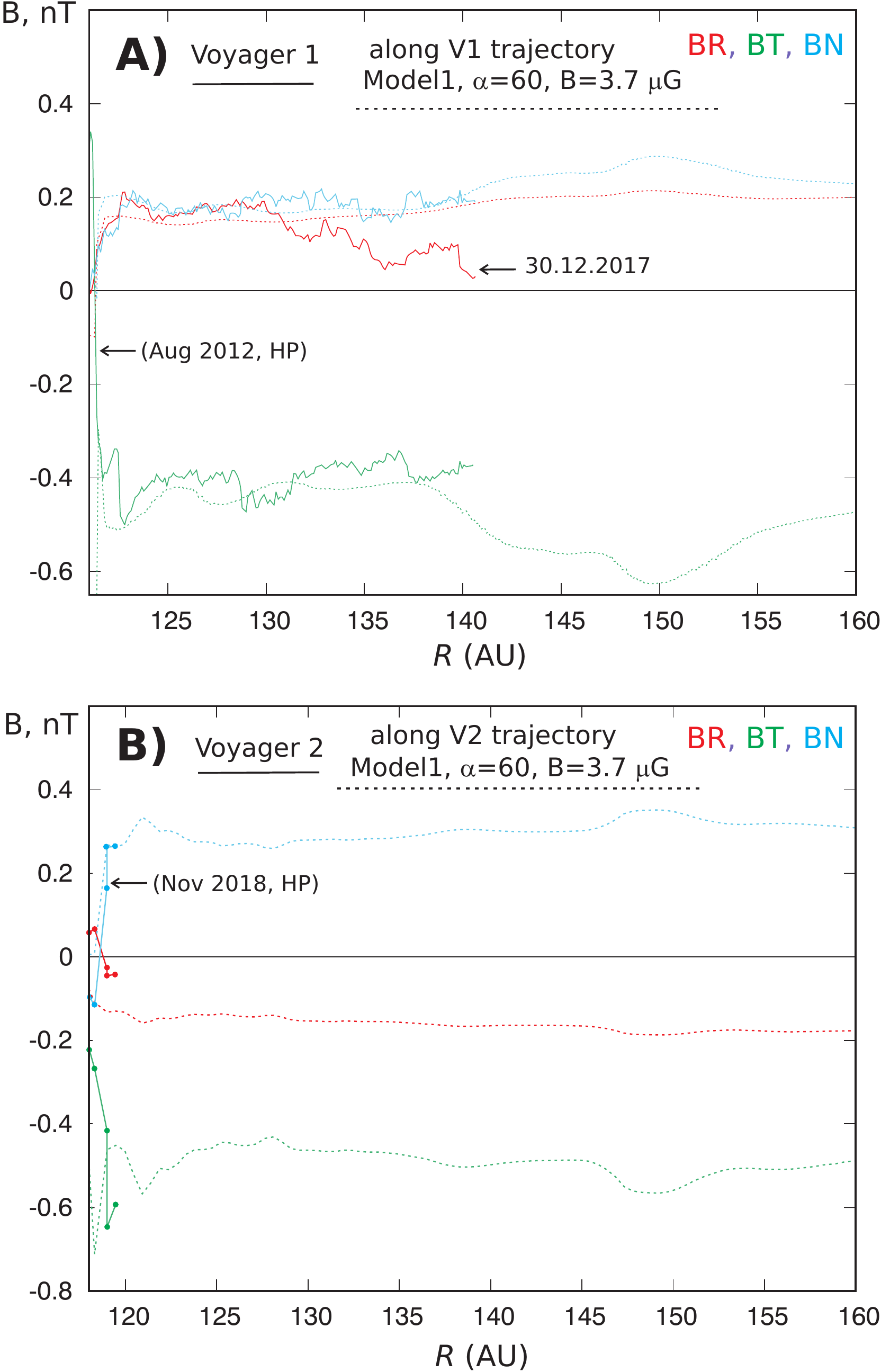}
        \caption{Components of the magnetic field (panels A and B ) in the RTN coordinate system along the Voyager 1 (panel A)) and Voyager 2 (panel B) trajectories. The results of Model 1 are shown as dashed curves. Solid curves represent Voyager 1/MAG data in Panel A and Voyager 2/MAG data in Panel B. The model curves are shifted by 3 AU in panel A, and by 5 AU in panel B.
        }\label{fig3}
\end{figure}

Figure \ref{fig3} presents the magnetic field and plasma velocity components along the Voyager 1 and 2 trajectories in the 
RTN coordinate system  as a function of time. Voyager data are shown as solid curves. 
Voyager 1 Magnetometer (MAG) data plotted in Figure \ref{fig3}A were obtained from the NASA website: https://cohoweb.gsfc.nasa.gov. These data were presented and discussed in \cite{burlaga_2018, burlaga_2019}.
Voyager 2/MAG data  were obtained from Fig. 3 of \cite{burlaga_2019b}.
Model results are shown as dashed curves.  Since the distances to the heliopause obtained in the model are larger than those measured from observations, we shifted the model distances by $\Delta R = R_{model} - R_{data}$  closer to the Sun and did not make any time shift. Here,   $\Delta R$ is 3 AU in Voyager 1 direction and 5 AU in Voyager 2.

The model--data comparison shows that just beyond the heliopause all three components of the interstellar magnetic field of  Model 1 match those measured by both Voyagers (Figs. \ref{fig3}A and \ref{fig3}B). Data for Voyager 2 are restricted by the closest vicinity of the crossing, while Voyager 1 magnetometer data after the heliopause crossing are publicly available from 2012 until the end of 2017. Therefore, we can compare how the three components of the magnetic field behave along the Voyager 1 trajectory.  Figure \ref{fig3}A shows that the $B_N$ and $B_T$ components match the data for the entire period of available observations. However, the small-scale details of the component fluctuations are different between the model and data; the model results are smooth. We associate these differences with the inner boundary conditions taken at 1 AU which are 27-day averages.

The most significant difference between Model 1 and the data exists for the $B_R$ component. 
In the Model, $B_R$ remains almost constant for the entire period from 2012 to 2017 and beyond, while in the data, the $B_R$ component is almost constant for 2013-2014 and then gradually decreases. Two bumps on the generally decreasing slope are associated with disturbances induced by merged interaction regions (\cite{burlaga_2018, burlaga_2019}). Possible reasons for and implications of the difference in the $B_R$ component  are discussed in the following section.


\section{Conclusions and Discussion}

 We have found a combination of magnitude and direction of the interstellar magnetic field ($B_{LISM}$=3.75 $\mu$G, $\alpha$ = 60$\circ$) for which the model reproduces the positions of the heliopause in Voyager 1 and 2 directions  reasonably well at the times of actual crossings. The difference of 2.5 AU is insignificant and can easily be removed by a slight increase of proton or H atom number densities. At the same time, the distance to the termination shock obtained  in the model is 84 AU in 2004 in Voyager 1 direction. This is 10 AU closer than the actual distance.  However, the model shows that in 2004-2005 the distance of the termination shock fluctuates rather strongly; this is connected with variations of the solar wind parameters. In reality, the variations are expected to be much stronger than in the model. Indeed, the angular resolution is 10 degrees in latitude and the time-resolution is 27 days in the model.  Stronger time variations of the termination shock can in principle reduce the model--data difference in the termination shock distance.  In Voyager 2 direction, the TS is 5.5 AU closer to the Sun in the model as compared to the actual crossing.

Therefore, the thickness of the inner heliosheath is $\sim$8 AU larger in the model. This difference is much smaller than for the model reported in Izmodenov and Alexashov (2015; referred to here as Model 2). Nevertheless, the difference is significant. We suppose that it may be due to some
yet unknown dissipation process in the inner heliosheath. The possible candidates were discussed in previous papers by \cite{Izmodenov_2014}  and \cite{izmod_alexash_2015}. 

The model reproduces all three magnetic field components just after the heliopause crossings by Voyager 1 and Voyager 2. The comparison of the model and data along the trajectory of Voyager 1 shows that  $B_T$ and $B_N$ components are predicted by the model relatively well. Radial component $B_R$ decreases in 2016-2017 in the data and remains constant in the model. 

It is important to note that the interstellar flow in the model is subsonic with respect to fast magnetosonic waves (i.e., fast magnetosonic Mach number is smaller than one). Therefore, there is no bow shock and, theoretically, interstellar flow is perturbed by the interaction with the Sun up to very large 
distances. However, from the numerical results (see Figure 1-B1) it is seen that at distances of $\approx$500 AU in the upwind direction the magnetic field lines are very weakly disturbed and the interstellar parameters  are close to those in the pristine LISM.

As discussed in Appendix B, in the inner heliosheath, the model reproduces $V_R$ and $V_T$ components in Voyager 1 direction relatively well (\ref{figB1}). In the direction of Voyager 2, the model reproduces $V_R$ and $V_T$ components just after the termination shock. However, at larger distances,  the components obtained in the model behave differently as compared to Voyager 2 plasma data. Apparently, the model reproduces the flow in Voyager 1 direction and does not reproduce it in Voyager 2 direction.  
As for the heliospheric magnetic field, the model shows a significant increase towards the heliopause that is not observed. The deficit of the magnetic field is probably due to a dissipation process. The magnetic field dissipation could also be the reason for the difference in the inner heliosheath plasma flow.

In the qualitative discussion of Section 2, we implicitly assumed that the shape of the heliopause is determined by the interstellar magnetic field. However, it has been shown  \citep{izmod_alexash_2015} that heliolatitudinal dependence of the solar wind dynamic pressure and heliospheric magnetic field influence the shape of the heliopause. In addition, but to a lesser extent, time variations may temporarily affect the heliopause shape locally. All these effects are taken into account in our model, and they are essential for the model results matching the data at the heliopause. means that 

The model--data difference in $B_R$ along the Voyager 1 trajectory may be connected with the time fluctuations. Indeed, when the heliopause moves in and out it acts as a piston; moving in, it compresses the magnetic field lines. Therefore, the tangential components of the interstellar magnetic field increase. The radial motion of the heliopause should increase the $B_R$ component  as well because the direction of the magnetic field should be parallel to the heliopause surface. The fact that fluctuations of all magnetic field components correlate is clearly seen in both Voyager 1 data and in the model results (Figure \ref{fig3}A).  However, the level of fluctuations in the model is somewhat smaller than in the data. This may be connected
with the dataset for the solar wind parameters that we employed in the model.
For example,  it may be due to the 27-day time averaging that was performed for the solar wind parameters in the model. Averaging indeed reduces the level of fluctuation. In addition, some instabilities may occur at the heliopause which are not considered in the model.

We also compared the results of the presented model with other constraints of the SW/LISM interaction model:
1) The magnetic field chosen for Model 1 
belongs to the hydrogen deflection plane established by \cite{Lallement_2010}; the angle between the direction of the interstellar hydrogen and helium in the outer heliosphere (for instance at 60-70 AU) is $\approx$3.5$^\circ$ which is similar to the deflection obtained by \cite{Lallement_2010}). 2) The interstellar hydrogen number density in the outer heliosphere is $\approx$0.1 cm$^{-3}$ which is similar to the number density established previously by different methods, e.g. by analyses of the supersonic solar wind deceleration \citep{Wang2003}, by analyses of Ulysses pickup protons data \citep{Gloeckler1997}, by analyses of SOHO/SWAN backscattered solar Lyman alpha data \citep{Quemerais2013}
 3) The direction of the magnetic field of Model 1  ($\approx$125$^\circ$, $\approx$37$^\circ$) in HGI is not too far from the direction derived by \citealp{Zirnstein_2016} from analyses of IBEX ribbon  (147.5$^\circ$, 31$^\circ$), although some difference exists in both direction and magnitude. We plan to analyze the IBEX ribbon using our model in the future.

\begin{figure}[t!]
        \includegraphics[width=0.5\textwidth,clip=]{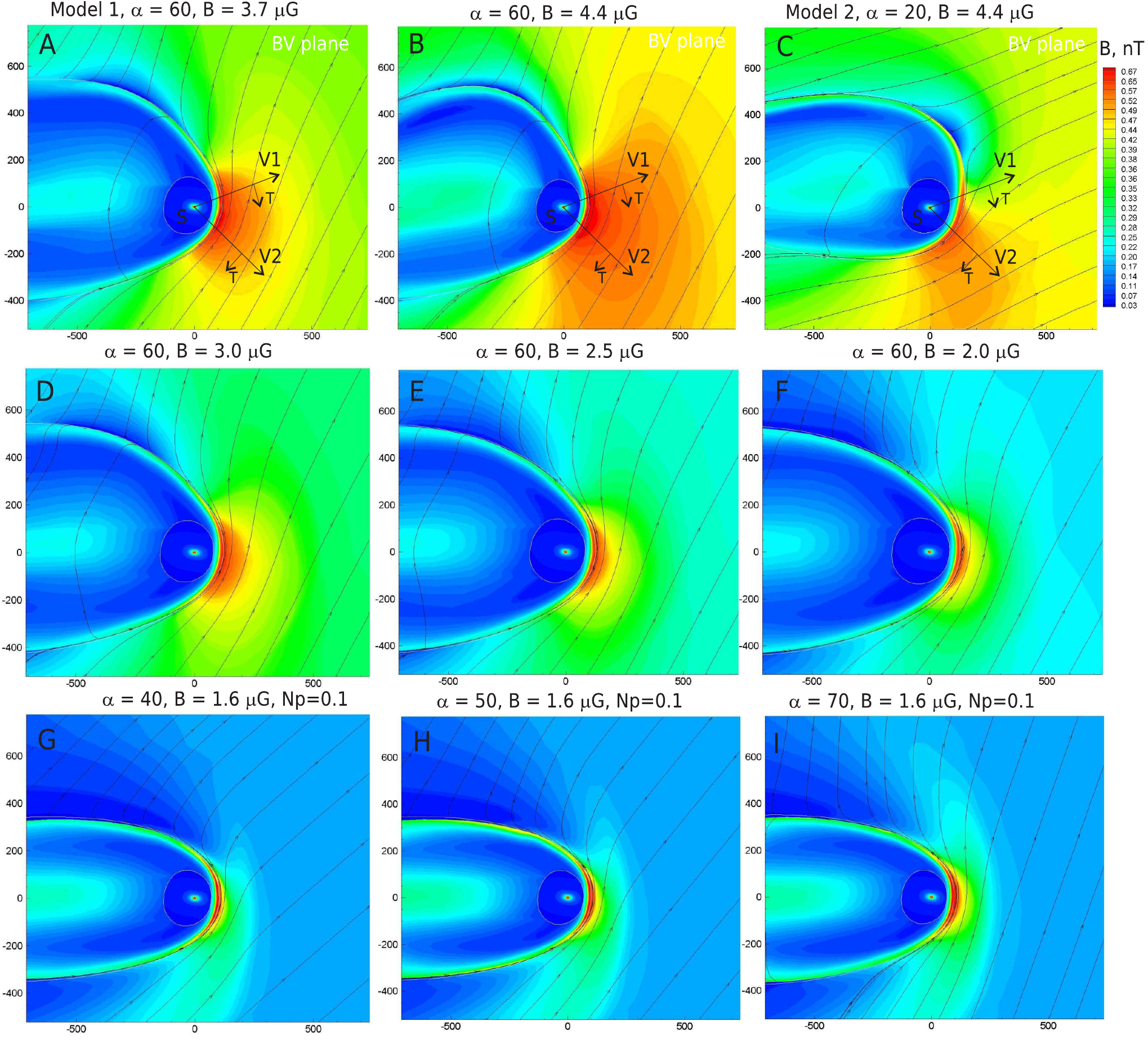}
        \caption{ Global shape of the heliopause, interstellar magnetic field lines, and magnitude in the outer heliosheath obtained for different model parameters.
        }\label{fig4}
\end{figure}

It is important to note that our numerical calculations were not restricted to only Models 1 and 2. We performed very intensive (although not completely systematic) parametric calculations by varying the magnitude of the interstellar magnetic field, the angle $\alpha$, and the interstellar proton and H atom number densities. Since our model requires us to solve 3D time-dependent MHD equations self-consistently with a 6D kinetic equation, such parametric calculations are extremely computationally expensive. This explains why we do not match the locations of the heliopause in Voyager 1 and Voyager 2 exactly. Such an "exact"\ result would be computationally expensive but of little use because of averaging of the solar wind parameters at the inner boundary. 
Nevertheless, Model 1 presented in this paper is the best model of many (about 50) considered models.
Figure \ref{fig4} presents (as an example) results of calculations with nine different model parameters. Model 1 is shown in panel A as a reference. The models with a larger interstellar field (one is shown in panel B) were ruled out because they produce a larger magnetic field at the heliopause as compared with Voyagers as well as smaller distances to the heliopause and especially to the termination shock.
The models with reduced magnitude of the magnetic field (panels D, E, F) produce a smaller magnetic field at the heliopause in Voyager 2 direction and opposite (as compared with Voyagers) asymmetry of the heliopause in Voyager 1 and 2 directions. Attempts to reduce the angle $\alpha$ (panel C) lead to a strong increase of the heliocentric distances to the heliopause.
In fact, it is also possible to produce the exact magnitude of the magnetic field at the heliopause in the models with a rather small magnetic field in the pristine LISM. To do that, we had to increase the dynamic pressure of the plasma component by increasing the interstellar proton number density (panels G, H,I). However, in this case, the heliopause became quite axis-symmetric and heliocentric distance to the heliopause became significantly larger in Voyager 2 direction as compared with Voyager 1. Therefore, such models were ruled out as well.

Here, we demonstrate qualitatively that the information on the interstellar magnetic field in Voyager 1 and 2 directions provides us with strong constraints on the direction and magnitude of the pristine interstellar magnetic field. Numerical results demonstrate that Model 1 matches the data relatively well but not completely. Further numerical studies will perhaps allow a model to be found that will fit the data better, but the resulting magnitude and direction of the interstellar magnetic field should not be too different from those presented here.

\section{Acknowledgements}
This work has been supported by Russian Science Foundation grant 19-12-00383. This paper profits from the discussions appeared at the workshop on the outer heliosphere that has been hold at Center of Advanced Studies of Russian Academy of Sciences on May 27-31, 2019.

\begin{appendix}

\section{Model and boundary conditions}
        
In this appendix we give a brief summary of the basic features of the Izmodenov and Alexashov (2015) model and discuss the boundary conditions used in time-dependent calculations presented here. The detailed formulation of the physical and numerical model is given in Izmodenov and Alexashov (2015).  In the model, partially ionized interstellar plasma is considered  as a two-component gas.
The two components are (1) a neutral component consisting of atomic hydrogen and (2) a charged (or plasma) component consisting of protons, electrons, and helium ions. In the solar wind, the plasma component consists of protons, alpha particles, and electrons.
The neutral component is described kinetically. The charged (plasma) component is described in the context of an ideal MHD approach. The influence of charge exchange with the interstellar H atoms was taken into account in the right-hand side of the MHD equations by the source terms. The sources are calculated as integrals of the H-atom velocity distribution function. Source terms have been calculated using the  Monte Carlo method (Malama 1991) which was used to solve the kinetic equation as well (see, Izmodenov et al, 2001). This method assumes that the velocity distribution function of the proton component is locally Maxwell, although it allows a generalization to any isotropic distribution function as was done by Malama et al. (2006). 

The inner boundary for our model is located at 1 AU. We take into account  helio-latitudinal variations of the SW density and speed at 1 AU obtained by using three different sets of data:
\begin{enumerate}
        \item In the ecliptic plane, we use data (SW density and speed) from the OMNI 2 dataset (https://omniweb.gsfc.nasa.gov/). The OMNI 2 data set contains hourly resolution SW magnetic field and plasma data from many spacecraft in geocentric orbit and at the L1 Lagrange point.
        \item Heliolatitudinal variations of the SW speed are taken from analysis of the interplanetary scintillation (IPS) data (Tokumaru et al 2012; http://stsw1.isee.nagoya-u.ac.jp/annual$\underline{\hspace{0.2cm}}$map.html).  Data are available from 1990 to 2017.
        \item Heliolatitudinal variations of the SW mass flux are derived from analysis of {\it SOHO}/SWAN full-sky maps of the backscattered Lyman-alpha
        intensities (Qu\'{e}merais et al. 2006, Lallement et al. 2010, Katushkina et al. 2013, 2019). An inversion procedure (see Qu\'{e}merais et al. 2006 for details)
        allows us to obtain the SW mass flux as a function of time and heliolatitude with a temporal resolution of approximately 1 day and an angular resolution
        of 10$^{\circ}$. Data are available from 1995 to the end of 2017.
\end{enumerate}

From these data, we calculate the mass and momentum fluxes of the SW as functions of time (from 1995 to 2017) and heliolatitude. 
We note that a similar procedure was followed to obtain the boundary conditions in Izmodenov and Alexashov (2015) with further averaging of the parameters over the entire time period. The distribution of the solar wind proton number density and velocity at 1 AU as functions of time and heliolatitude are shown in Figure A.1. The solar wind dynamic pressure is shown in Figure A.1 (plot C) as well. The heliospheric magnetic field is described exactly as it was in Izmodenov and Alexashov (2015), so we do not repeat it here.

The interstellar bulk velocity and temperature are chosen as V$_{LISM}$ = 26.4 km s$^{-1}$ and T$_{LISM}$ = 6530 K. This velocity value is consistent with the ISM vector derived from analyses of {\it Ulysses}/GAS data for interstellar helium (Witte 2004, Bzowski et al. 2014, Katushkina et al. 2014, Wood et al. 2015).
The direction ${\bf V}_{LISM}$ is determined by the interstellar helium flow direction from {\it Ulysses}/GAS data. This value is (longitude=-1.02$^{\circ}$, latitude=-5.11$^{\circ}$) in the heliographic inertial coordinate system (HGI 2000).
The velocity vector is in agreement with new results derived from {\it IBEX} data (see, e.g. McComas et al. 2015).

The interstellar H atom and proton number densities are assumed to be the same as in Izmodenov and Alexashov (2015): n$_{H, LISM}$ = 0.14 cm$^{-3}$ and n$_{p, LISM}$ = 0.04 cm$^{-3}$, respectively. The number density of the interstellar helium ions that provide additional pressure (see Izmodenov et al., Izmodenov and Alexashov, 2015) is chosen as $n_{He+, LISM} = 0.003$ cm$^{-3}$.

The remaining parameter is the IsMF. As in Izmodenov and Alexashov (2015) we assume that the plane which is determined by the vectors ${\bf V}_{LISM}$  and ${\bf B}_{LISM}$  ({\it BV}-plane) coincides with the plane determined by the velocity vectors of the interstellar helium, ${\mathbf V}_{He}$,  and hydrogen, ${\mathbf V}_{H}$, inside the heliosphere -- the so-called hydrogen deflection plane (HDP; see, Lallement et al. 2005, 2010; for model justification see Izmodenov et al., 2005b, 2009). 
Under this assumption, the IsMF is determined by its magnitude and by the angle  with the direction of the interstellar bulk flow, ${\mathbf V}_{He, LISM}$. This angle is denoted as $\alpha$. Izmodenov and Alexashov (2015) assumed B$_{LISM}$ = 4.4 $\mu$ Gauss and $\alpha$=20$^{\circ}$. The values were obtained in the parametric study presented in Table 2 of Izmodenov et al. (2009) to obtain asymmetry of the TS in agreement with the distances of the {\it Voyager 1} and {\it Voyager 2} TS crossings. However, as was shown by Izmodenov and Alexashov (2015), our model with such a IsMF obtains the heliopause at very large distances, which are in disagreement with the actual Voyager crossings of the heliopause.

\begin{figure*}[t!]
        \includegraphics[width=\textwidth,clip=]{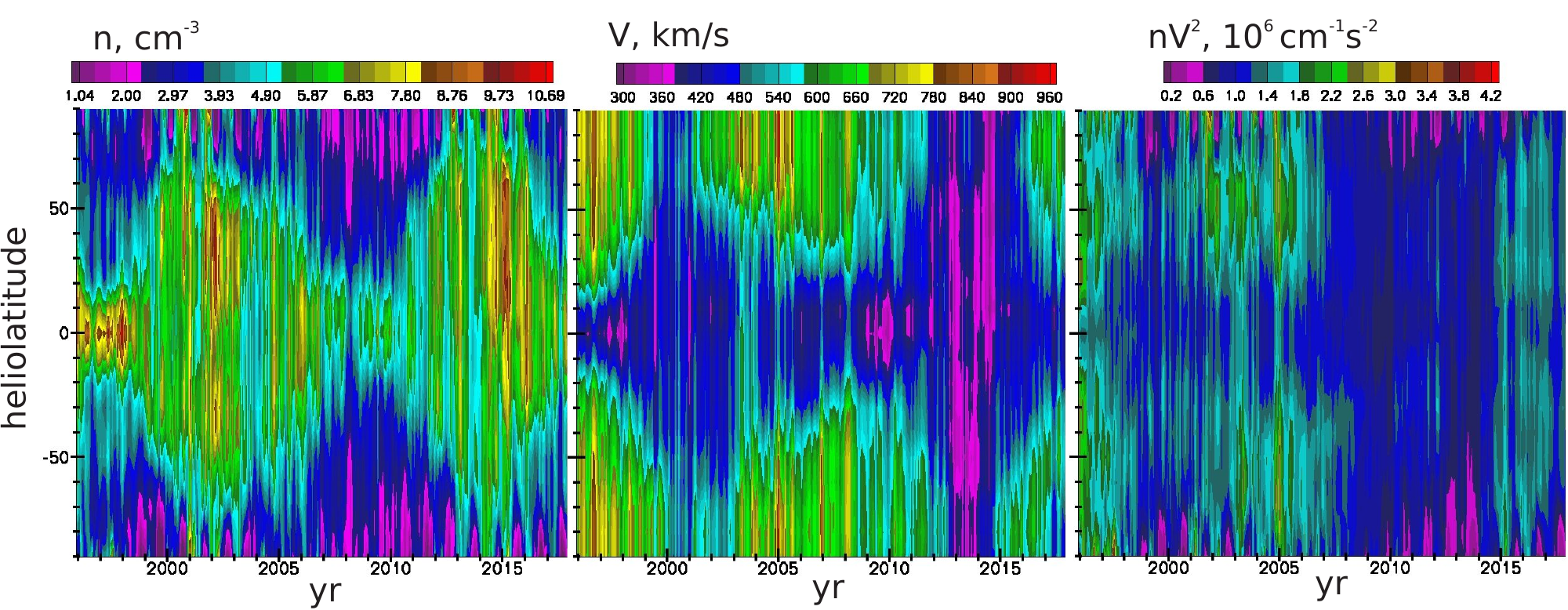}
        \caption{Solar wind proton number density (plot A) and velocity (plot B) at 1 AU and dynamic pressure (plot C) as functions of time and heliolatitude.}\label{sw_1AU}
\end{figure*}

\section{Comparison of the model with Voyager data in the inner heliosheath}

\begin{figure}[t!]
        \includegraphics[width=0.5\textwidth,clip=]{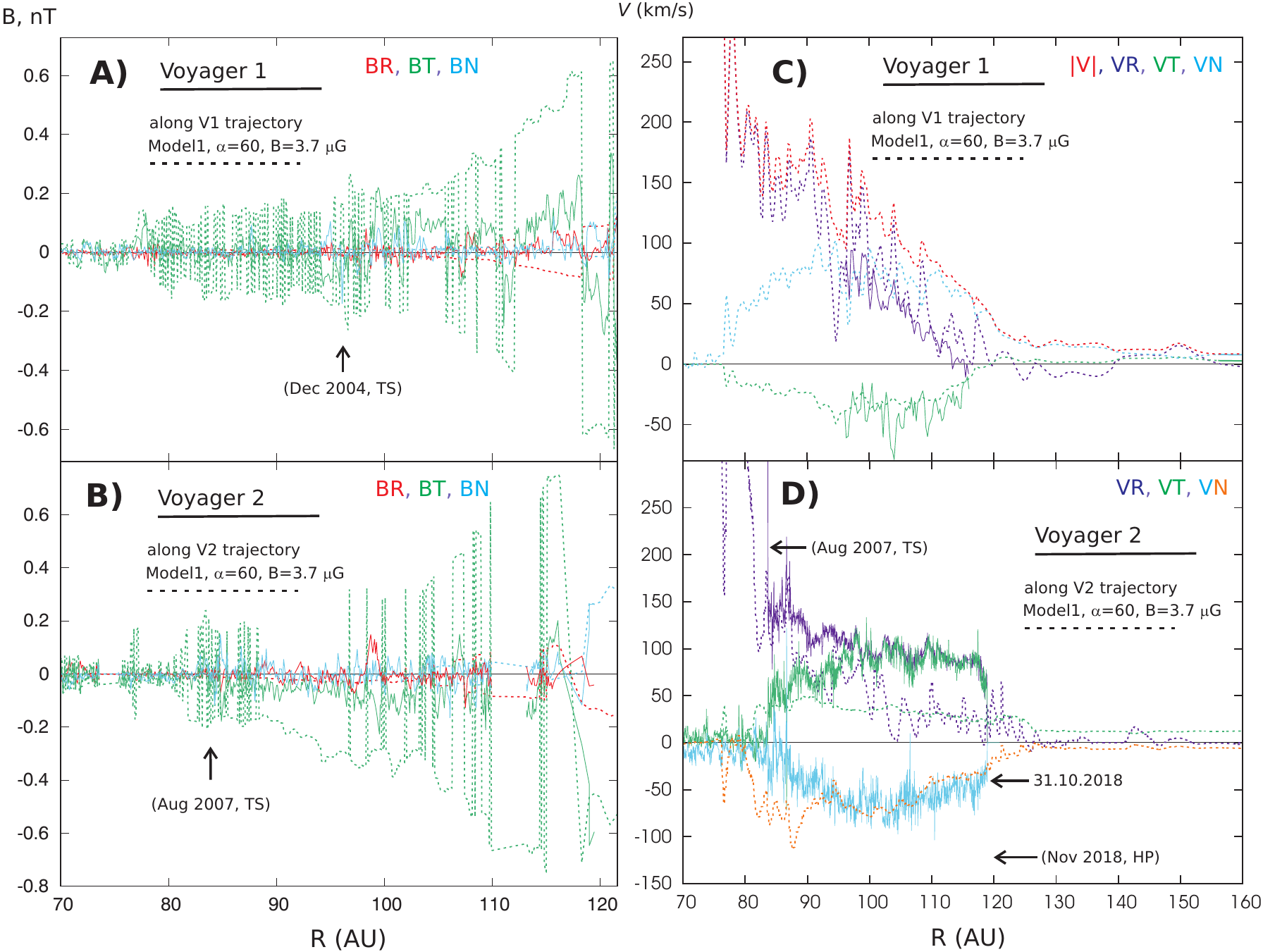}
        \caption{Components of the magnetic field (panels A and B ) and plasma velocity (panels C and D) in RTN coordinate system along the Voyager 1 (top panels) and Voyager 2 (bottom panels) trajectories in the inner heliosheath. Results of Model 1 are shown as dashed curves. Solid curves represent Voyager 1/MAG data in Panel A, Voyager 2/MAG data in Panel B. $V_r$ and $V_T$ derived by \cite{Krimigis_2011} from analyses of energetic particle anisotropy measured by Voyager 1/LECP are shown as dots in panel C. Dots in panel D show Voyager 2/PLS  data. 
        }\label{figB1}
\end{figure}

Although comparison of the model results with Voyager data from the inner heliosheath (i.e., the region between the termination shock and the heliopause) is not the major purpose of this paper, it is presented in Figure \ref{figB1} as well. 
Figure \ref{figB1} (panel C) shows $V_R$ and $V_T$ plasma velocity components obtained in Model 1 with those derived from analyses of anisotropy of energetic particle fluxes \citep{Krimigis_2011}. Relatively good qualitative agreement of model and data was obtained. The $V_R$ component is about $\sim$100 km/s after the  decreases and becomes zero or sometimes negative as the spacecraft approaches the stagnation region (Krimigis et al., 2011). The $V_T$ component is about 30-40 km/s at the TS and slowly decreases toward the stagnation region.

Comparison of the model results with Voyager 2/ PLS data is shown in Figure \ref{figB1} (panel D). Up to now these data were only available  for the period up to October 1, 2018, that is before Voyager 2 crossed the heliopause.
First of all, although not shown in the figure, we notice that the model reproduces the speed and its fluctuations in the supersonic solar wind reasonably well. 
The location of the TS is closer in the model by $\sim$5 AU in Voyager 2 direction. Just after the TS, the magnitudes of the $V_R$ and $V_T$ components of velocity coincide with the data remarkably well.
As the spacecraft moves out, the $V_R$ component decreases both in the model and in the data. However, the decrease obtained in the model is much faster.  
The $V_R$ component behaves in Voyager 2 direction in the similar way as it was observed earlier in the direction of Voyager 1.
 This is not observed in Voyager 2/PLS data. The
$V_T$ component remains almost constant (or slowly decreases)  in the model, while  in the data this component increases from $\sim$80 km/s just after the TS crossing to $\sim$100 km/s. The
$V_N$ component is larger in the model immediately after the shock. It is interesting enough that for the distances larger than 90 AU the model reproduces V2 observations for this component relatively well.

Finally, we compare the magnetic field in the inner heliosheath (Figures \ref{figB1}, panel A and B).
The magnetic field (mostly $B_T$ component) obtained in the model can be seen to significantly increase towards the heliopause. This is the magnetic pile-up region that was discussed in \cite{izmod_alexash_2015} in detail. It is important to note that this effect is obtained in the frame of an ideal MHD approach. The fact that it is not observed may suggest that there is effective dissipation of the magnetic field in this region.

\end{appendix}

\end{document}